\documentclass[iop]{emulateapj}

\newcommand{\myemail}{maggie@physics.mcgill.ca}
\newcommand{\psr}{PSR~J1846$-$0258}
\newcommand{\kes}{PSR~J1846$-$0258}
\newcommand{\snr}{Kes 75}
\newcommand{\rxte}{{\textit{RXTE}}}

\newcommand{\nudotdotdot}{{\ifmmode\stackrel{\bf \,...}{\textstyle \nu}\else$\stackrel{\,\...}{textstyle \nu}$\fi}}
\newcommand{\degrees}{^{\circ}}

\newcommand{\xte}{{\it RXTE}}

\usepackage{graphicx}

\shorttitle{Post-Outburst Observations of \psr}
\shortauthors{Livingstone et al.}

\begin{document}

\title{Post-outburst Observations of the Magnetically Active Pulsar
J1846$-$0258. a new braking index,
increased timing noise, and radiative recovery}

\author{Margaret A.~Livingstone \altaffilmark{1},
C.-Y.~Ng\altaffilmark{2}, Victoria M.~Kaspi}
\affil{Department of Physics, Rutherford Physics Building, 
McGill University,  Montreal, QC H3A 2T8, Canada}

\author{Fotis~P.~Gavriil}
\affil{NASA Goddard Space Flight Center, Astrophysics Science Division, Code
662, Greenbelt, MD 20771, USA}
\affil{Center for Research and Exploration in Space Science and Technology,
University of Maryland Baltimore County, 1000 Hilltop Circle,
Baltimore, MD 21250, USA}

\and
\author{E.~V.~Gotthelf}
\affil{Columbia Astrophysics Laboratory, 550 West 120th Street, 
Columbia University, New York, NY 10027-6601, USA}

\altaffiltext{1}{\myemail}
\altaffiltext{2}{Tomlinson Postdoctoral Fellow}

\begin{abstract}
The $\sim$800-year old pulsar J1846$-$0258 is a unique transition object
between rotation-powered pulsars and magnetars: though behaving 
like a rotation-powered pulsar most of the time, in 2006 it exhibited 
a distinctly magnetar-like outburst accompanied by a large glitch.  
Here we present X-ray
timing observations taken with the {\textit{Rossi X-ray Timing
Explorer}} over a 2.2 year period after the X-ray outburst and 
glitch had recovered. We observe that the braking index of the pulsar,
previously measured to be $n=2.65\pm0.01$, is now $n=2.16\pm0.13$, a
decrease of $18\%\pm5\%$. We also note a persistent increase in the
timing noise relative to the pre-outburst level.
Despite the timing changes, 
a 2009 {\textit{Chandra X-ray Observatory}} observation shows that
the X-ray flux and spectrum of the pulsar and its wind 
nebula are consistent with
the quiescent levels observed in 2000. 
\end{abstract}


\section{Introduction}
\label{sec:intro}

\objectname[PSR J1846-0258]{\kes} is the very young ($\sim$800~year) X-ray 
pulsar at the center of the
supernova remnant \objectname{Kes 75} \citep{gvb+00}.
The pulsar has a spin period of 326-ms and usually exhibits properties common
to those of rotation-powered pulsars (RPPs), including powering a 
bright pulsar wind nebula (PWN).
This pulsar is notable for having an unusually high spin-down-inferred
magnetic field ($B=5 \times 10^{13}$\,G), and is one of the few with a
measured braking index
\citep[$n \equiv \nu \ddot\nu/{\dot\nu}^2 = 2.65\pm0.01$, where $\nu$
is the spin-frequency, and $\dot\nu$ and $\ddot\nu$ are its
derivatives;][]{lkgk06}. 
Measured braking indices fall in the range $1.4<n<2.84$
\citep{lps93,lpgc96,lkg05,lkgm05,lkgk06,wje10}, 
all less than the canonical value of $n=3$ predicted for magnetic dipole
radiation in a vacuum \citep[e.g.,][]{go69}. Explanations for $n<3$ include 
an increasing
magnetic moment \citep[e.g.,][]{br88,lyn04} or the effects of
magnetospheric plasma on the spin-down torque \citep[e.g.,][]{hck99,klo+06}.

Against expectations, \psr\ exhibited distinctly
magnetar-like behavior in 2006 May-July, when it showed several X-ray bursts,
an X-ray flux increase \citep{ggg+08}, a sizable rotational glitch with
remarkable ``overshoot'' recovery \citep{lkg10,kh09}, as well as spectral 
changes \citep{ks08,ggg+08,nsgh08}. 
This pulsar, evidently an RPP/magnetar transition object, 
presents a unique opportunity of 
exploring the long-term relationship between magnetic activity and
neutron star spin-down. 

Bursts of X-rays and variable X-ray flux, as
observed in most magnetars and \kes, are proposed to originate from
small- or large-scale reorganizations of the magnetic field
\citep[e.g.,][]{tlk02}. 
As the magnetic field is intimately 
connected with the temporal evolution of pulsars, comparing
timing behavior before and after an episode of magnetic activity
in a neutron star could provide important insight into the physics 
of neutron star magnetospheres.
None of the magnetars have  measured braking indices, despite their young
ages ($\tau_c \sim 1-100$\,kyr)\footnote{See
http://www.physics.mcgill.ca/$\sim$pulsar/magnetar/main.html
for a catalog of known magnetars.}. This is because of 
a significant level
of low-frequency timing noise \citep{gk04} and in several cases, large glitches
\citep[][and references therein]{dkg08a}, which prevent a measurement
of $n$ in every known magnetar. 
By contrast, \kes\ rotates relatively steadily, allowing for a
measurement of $n$ with regular monitoring observations with the
{\textit{Rossi X-ray Timing Explorer}} (\rxte). 
Between its discovery in 1999 \citep{gvb+00} 
and the outburst and glitch in 2006, it experienced 
a level of timing noise typical of young RPPs along with 
one small glitch ($\Delta\nu/\nu \sim 2.5 \times
10^{-9}$ near MJD~52210) and one small candidate glitch
($\Delta\nu/\nu < 5 \times 10^{-8}$ near MJD~52910).

Also available from regular \rxte\ monitoring is the pulsed flux of \kes.
As reported by \citet{ggg+08},
the pulsed flux had returned to its quiescent value roughly two
months after the 2006 May outburst and no further flux
variations have been observed \citep{lkg10}. However, because \rxte\ is a
non-focusing X-ray telescope, no information about the total flux or the
phase-averaged spectrum are available from these data. 
To measure these quantities and to confirm the \rxte\ results, 
a focusing X-ray instrument like the {\textit{Chandra X-ray Observatory}}
is required.  

Indeed, {\it Chandra} observations of the pulsar and PWN revealed
flux and spectral changes at the time of the 2006 outburst
\citep{ggg+08,ks08,nsgh08}. 
The pulsar's total flux rose considerably, but equally as interesting
was its change in spectrum.  The quiescent spectrum of \psr\ is much
like that of other young, high spin-down luminosity RPPs: a simple 
power law.  {\it Chandra} observations 
showed that while in outburst, the spectrum softened significantly
such that it became reminiscent of those observed from Anomalous X-ray
Pulsars (AXPs), namely, well described by a power law with an additional
thermal component. In addition, the superior 
angular resolution of {\emph{Chandra}} allowed detailed observations
of the PWN, showing that a marginal 
increase in flux may have occurred
between 2000 and immediately post-outburst \citep{ks08}. 
The effect of magnetar-like
outbursts on PWNe is an open question, as none of the bona fide
magnetars power nebulae 
(although see \citet{vb09} and \citet{tve+10} for discussion of a
possible PWN surrounding the AXP~1E~1547.0$-$5408). 
\citet{ks08} suggested a causal relation between a possible increase
in the PWN flux observed in 2006 and this or past magnetar-like
outbursts of the pulsar, 
while \citet{kp08}
suggested that the PWN was over-luminous compared to those of 
other young pulsars, perhaps owing to previous, unseen magnetar outbursts. 
However, a revised distance estimate of $\sim$6-kpc \citep{lt08}, 
rather than the previously claimed 
19\,kpc \citep{bh84}, reduces the implied nebular X-ray
efficiency to $\eta = L_{\rm{PWN}}/{\dot{E}} \simeq 0.02$. While this is
still large, it is similar to that observed from the Crab, hence need
not be powered by previous outbursts. 
Thus any long-term effect of the 2006 outburst on the PWN could help
clarify this point. 

In this paper, we report on 2.2 year of \rxte\ timing
observations of \kes\ in the post-magnetic activity, post-glitch
recovery period. We perform phase-coherent and partially phase-coherent
timing analyses after the 2006 glitch had largely recovered, 
and report a post-burst measurement of $n=2.16\pm0.13$,
smaller than the pre-outburst value at the 3.8$\sigma$ level.
We also quantify the increase of the timing noise observed over the bursting
episode and discuss the implications of these observations. In
addition, we report on 2009 \emph{Chandra}
observations of the pulsar and associated PWN which
show that the pulsar and PWN flux and spectra are consistent with their
pre-outburst values. 

\section{\rxte\ Observations and Analysis}
\label{sec:obs} 
{\textit{RXTE}} observations of \psr\ were made using the Proportional Counter Array 
\citep[PCA;][]{jsg+96,jmr+06}. The PCA consists of an
array of five collimated xenon/methane multi-anode proportional counter
units (PCUs) operating in the 2~--~60\,keV range, with a total effective
area of approximately $\rm{6500~cm^2}$ and a field of view of 
$\rm{\sim 1^o}$~FWHM. 

The \rxte\ data of \psr\ are unevenly spaced over 
11\,years from 1999 April 18 through
2010 April 22 (MJD 51286~--~55308). Observations taken between 
1999 April 18 and 21 (MJD 51286~--~51289) 
are excluded because they cannot be unambiguously phase connected 
with the rest of the data. 
Data from 2000 January 31 to 2008 December 10
(MJD 51574~--~54810) were reduced and analyzed
previously and details can be found in \citet{lkgk06} and
\citet{lkg10}. Data taken between 2009 January 27 and 2010 April 22 
(MJD 54858~--~55308) are described here for the first time.

Data were collected 
in ``GoodXenon'' mode, which records the arrival time
(with 1-$\mu$s resolution) and energy (256 channel resolution) of every
unrejected event. Typically, two to three PCUs were operational
during an observation. We used the first Xenon layer of each
operational PCU and extracted events in channels 4~--~48
(approximately 2~--~20\,keV), 
as this produces good quality profiles for individual observations. 

Observations were downloaded from the HEASARC 
archive\footnote{http://heasarc.gsfc.nasa.gov/docs/archive.html}
and data from each active PCU were merged and binned at
(1/1024)\,s resolution. Photon arrival times were converted
to barycentric dynamical time (TDB) at the solar system barycenter using the
J2000 source position R.A. = $18^{\rm{h}}46^{\rm{m}}24\fs94\pm0\fs01$,
decl. $= -02 \degrees 58\arcmin30.1\arcsec\pm0.2\arcsec$ \citep{hcg03}
and the JPL DE200 solar system ephemeris. 

The phase-coherent ephemeris from \citet{lkg10} was used to fold
the time series for each new observation 
with 16 phase bins. Resulting profiles were cross-correlated 
with a high significance template profile, producing
a single Time-Of-Arrival (TOA) for each observation. 
This process produced 265 TOAs between MJD 51574 and 55308 with a
typical uncertainty of $\sim 10$\,ms ($\sim$3\% of the pulse period).
The TOAs were fitted to a timing model using the pulsar timing
software package              
TEMPO\footnote{http://www.atnf.csiro.au/research/pulsar/tempo/}.
Further details of the timing analysis are given in \citet{lkgk06}.

\section{Timing Analysis and Results}

\subsection{Phase-coherent Timing Analysis}
In order to make a significant measurement of
a deterministic value of $\ddot\nu$ and thus $n$,
we restricted our phase-coherent timing analysis to
MJD~54492~--~55308 (2008 January 27~--~2010 April 22),
because earlier observations
are highly contaminated by glitch recovery and
timing noise, as discussed in \citet{lkg10}. 

\begin{figure}
\plotone{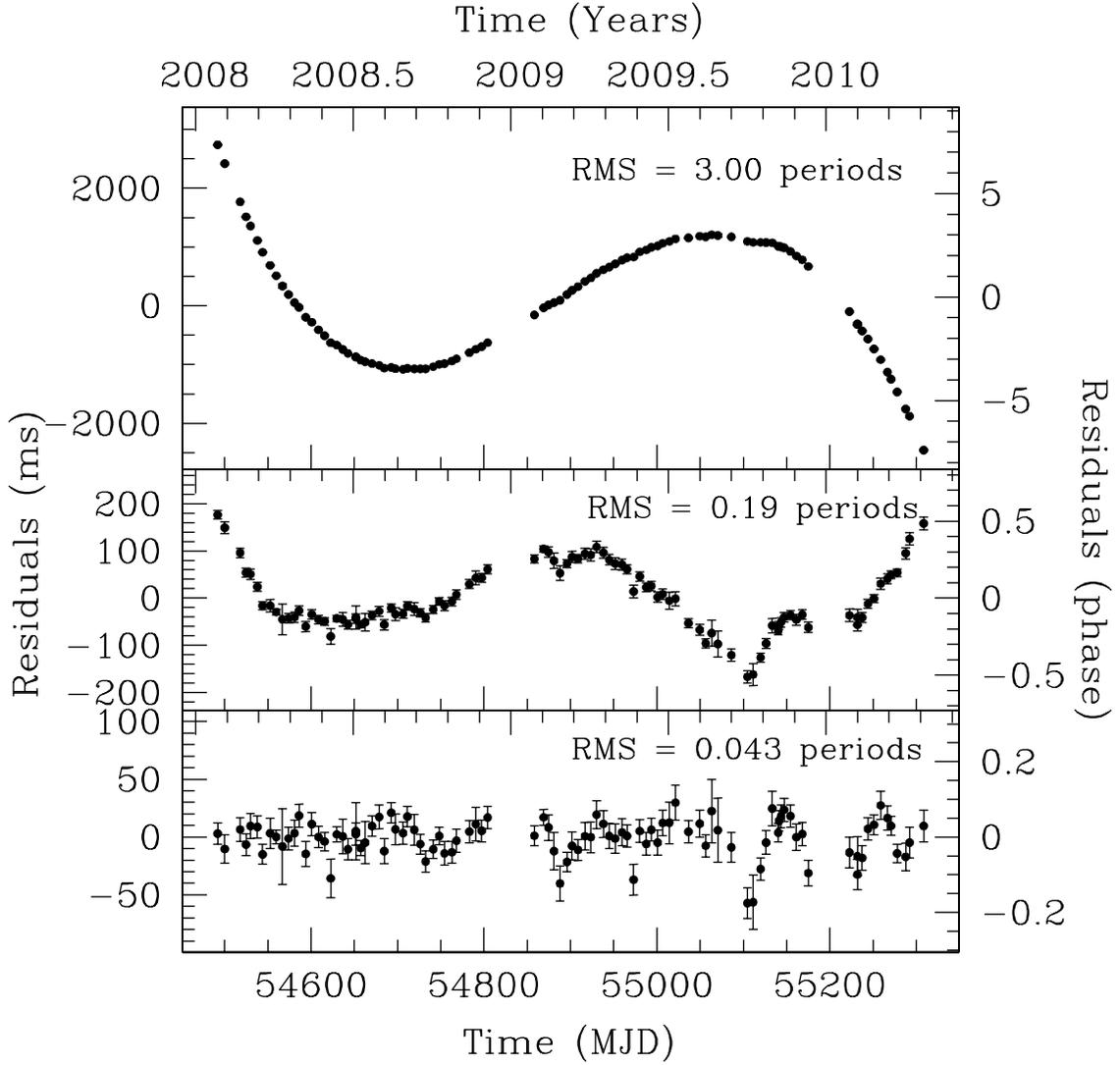}
\figcaption[Timing residuals of \kes\ spanning MJDs~54492~--~55308]
{\footnotesize{Timing residuals of \kes\ spanning MJDs~54492~--~55308
(2008 January 27 ~--~2010 April 22). The top panel shows residuals with 
$\nu$ and $\dot\nu$ fitted. The middle panel shows residuals with $\nu$, 
$\dot\nu$, and $\ddot\nu$ fitted, while the bottom panel shows residuals 
with 12 frequency derivatives fitted.\label{fig:kes_resids_2008_10}}}
\end{figure}

We obtained a single phase-coherent timing solution (with no phase
ambiguities) fitting only $\nu$ and $\dot\nu$ for the 100 TOAs during 
this time
period, shown in the top panel of
Figure~\ref{fig:kes_resids_2008_10}. The residuals show a
very significant phase contribution from a second frequency
derivative (i.e., $\sim$15 phase turns over roughly 2 years).
We therefore added $\ddot\nu$ to our phase-coherent fit,
and the resulting residuals are shown in the middle panel of the figure.
The fitted $\ddot\nu$ corresponds to a braking index
of $n=1.888\pm0.002$. 
As visible in the middle panel of Figure~\ref{fig:kes_resids_2008_10}, 
significant timing noise remains in the
data; the timing residuals are not Gaussian distributed.
As a result, the formal 1$\sigma$ uncertainty on $n$ from 
this global phase-coherent
fit underestimates the true uncertainty. 
Spin parameters from this fit 
are given in Table~\ref{table:kes75_3_parameters}.

We fitted higher order frequency derivatives to ``whiten'' the phase
residuals, a common procedure to lessen the contaminating
effect of timing noise on fitted parameters \citep[e.g.,][]{kms+94}.
Fitting 12 frequency derivatives (the maximum possible given the current
machine precision) removes the majority of the timing noise, 
though does not render the phase residuals entirely Gaussian,
as shown in the bottom panel of Figure~\ref{fig:kes_resids_2008_10}.
Table~\ref{table:n_v_der} shows the variation of $n$ as
derivatives are fitted, as well as ${\chi^2}_\nu$. The value of 
$n$ varies between 1.89 and 2.95 as
higher order derivatives are fitted, without converging to a single
value, rendering the true value of $n$ ambiguous from this analysis.
Nevertheless, the range of measured $n$ values from this analysis
is relatively narrow: the timing noise has increased and clearly
contaminates, but does not completely dominate $\ddot\nu$. In 
cases where a parameter is dominated by a noise process, 
it can be of several orders of magnitude
larger and often of the wrong sign \citep[e.g.,][]{hlk10},
neither of which are seen here. In cases where timing noise
contaminates a measurement of a deterministic parameter but may not
dominate, as for these \psr\ data, 
using a partially coherent timing analyis can be useful to
find the true value. 

\subsection{Partially Coherent Timing Analysis}
\begin{figure}
\plotone{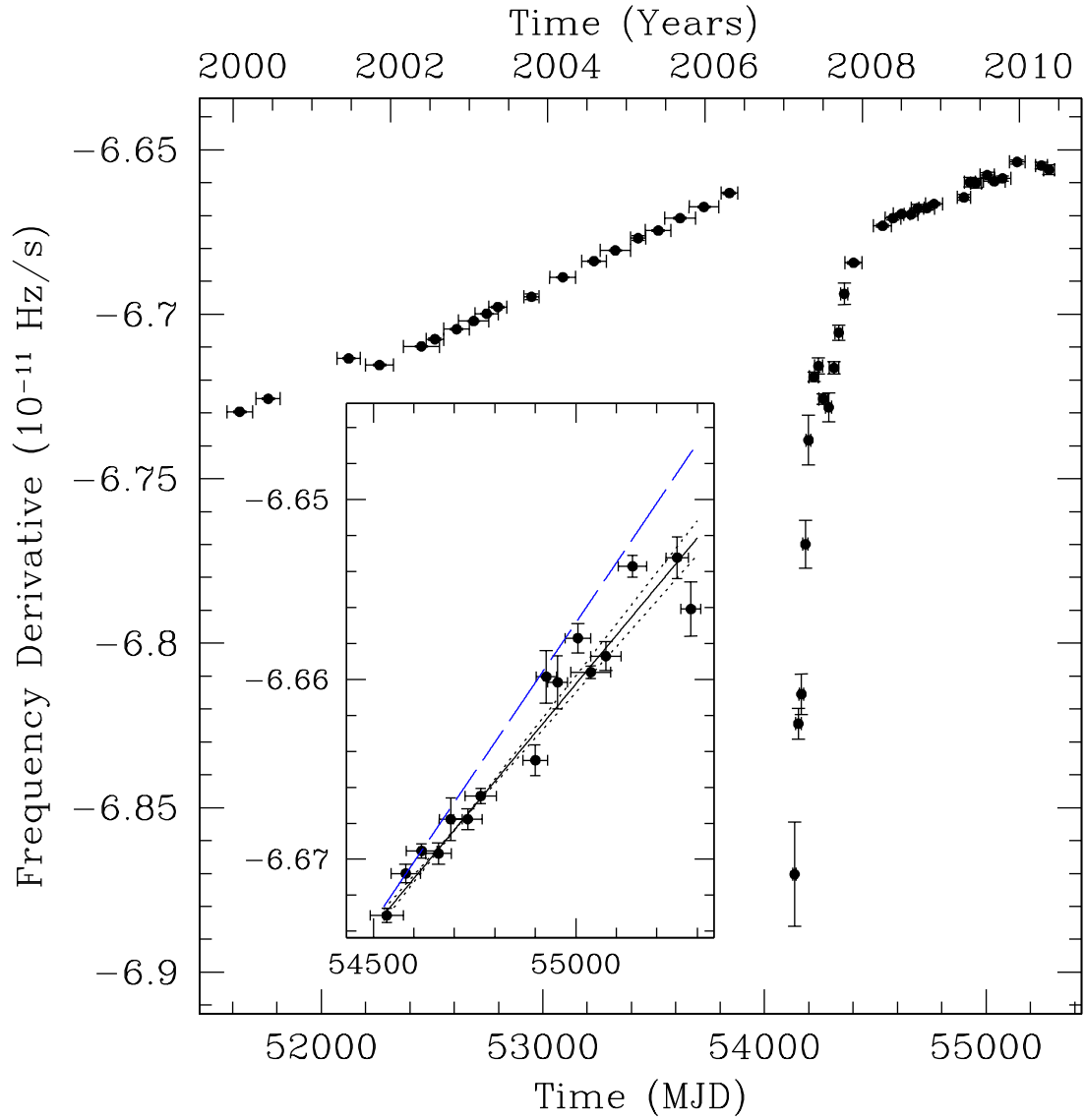}
\figcaption[Frequency derivative evolution of \kes\ over 
$\sim10$\,yrs of \rxte\ timing observations.]{\footnotesize{Evolution of the
frequency derivative of \kes\ over $\sim$10-years of \rxte\ timing
observations. Measurements of $\dot\nu$ overlap by $\sim$1/2 where sufficient data
are available. 
The effect of the two confirmed glitches is visible in the figure,
the first near MJD~52210 as a small discrete jump in $\dot\nu$ and the
second as the large decrease in $\dot\nu$ near MJD~53886 (note that
the full effect of the glitch on $\dot\nu$ is not shown here in the
interest of making visible the smaller changes in $\dot\nu$ at other
epochs, see \citet{lkg10} for additional details about the large
glitch). The inset shows measurements of $\dot\nu$
from MJDs~54492~--~55308, with the best-fit slope shown
as a solid line and the $\pm$1$\sigma$ uncertainties
(from the bootstrap analysis which better accounts for the increase in
timing noise) shown as dotted lines.
The slope corresponding to the pre-outburst $n$ is shown as
a dashed line (colored blue in the online version), where
uncertainties are roughly an order of magnitude smaller than those
from the post-outburst era, so are not visible in the figure.}
\label{fig:kes75_allnudot} }
\end{figure}

To mitigate the effects of timing noise, we performed a
partially coherent timing analysis. We created 48
short phase-coherent timing solutions spanning from 2000 to 2010
(MJDs 51574--55308). For each short timing solution, we fit only
$\nu$ and $\dot\nu$. 
The time span included in each subset was 
determined from the requirement that the reduced
$\chi^2$ of the fit was $\sim$1, and that no red noise-like
structure was visible in the data. 
In addition, we allowed the data included in each timing solution to
overlap by $\sim$1/2, where sufficiently dense sampling was available. This 
improves coverage, which is of particular importance while the timing
noise level is very high. 
Figure~\ref{fig:kes75_allnudot} shows the resulting $\dot\nu$
measurements spanning from 2000 to 2010. This analysis excludes 
observations taken between 
2006 May 31 and 2007 January 27, 
when glitch recovery and timing noise prevented a
coherent timing solution. 
As discussed in \citet{lkgk06} and confirmed by our current analysis,
from 2000~--~2006 May, $\dot\nu$ increased very
regularly, except at a small glitch in 2001.
The steady increase in $\dot\nu$ corresponds to a
braking index of $n=2.65\pm 0.01$, as measured from a phase-coherent
analysis of these data.
The large glitch (visible as a sudden decrease in
$\dot\nu$ by $\sim$3\% in Figure~\ref{fig:kes75_allnudot})
and the increase in timing noise 
had largely recovered by the beginning of 2008, as shown in the
figure. 
 
To obtain a post-burst measurement of $n$, 
we ignored all timing data prior
to MJD~54492 (2008 January 27), where the aforementioned 
glitch recovery and timing noise dominate. We performed a
weighted least-squares fit to 16 $\dot\nu$ measurements
spanning MJDs~54492~--~55308 (2008 January 27~--~2010 April 22, shown
in the inset of Figure~2). 
Given the large scatter in the post-burst $\dot\nu$ measurements, and
the known effects of timing noise on these data, 
it is likely that the formal uncertainties significantly
underestimate the true uncertainties. Thus, to better
estimate the uncertainty on $\ddot\nu$, we employed a 
bootstrap error analysis, helpful when formal
uncertainties may underestimate the true uncertainties
\citep{efr79}, and previously used for this same purpose
in \citet{lkg05,lkgm05}. This results in
$\ddot\nu = 3.13(19) \times 10^{-21}$\,s$^{-3}$,
corresponding to $n=2.16\pm 0.13$,
where the uncertainty from the bootstrap estimate
is larger than the formal uncertainty by a factor
of $\sim$2.4. This new measurement of $n$ is
smaller than the pre-outburst value of $n=2.65\pm0.01$ at the
3.8$\sigma$ level (or 9.1$\sigma$ 
if only the formal uncertainties are considered).
Thus, the braking index decreased by
$\Delta{n} = -0.49\pm 0.13$, following the period of
magnetar-like activity in 2006. This is the
first observed significant measurement of a
change of a braking index.

To further confirm a change in $n$, we performed an identical
partially coherent timing analysis
on the pre-outburst measurements of $\dot\nu$. Because of the small
glitch and candidate glitch, the data were separated into three large
subsets. The first subset (prior to the small glitch near MJD~52210)
contains only three measurements of $\dot\nu$, resulting in
$n=2.63\pm0.04$ (where uncertainties for this measurement are the
formal uncertainties because a bootstrap error analysis cannot be
performed with no additional degrees of freedom). 
 The second subset lies between the small glitch and a
78 day gap in the data (containing a candidate glitch).
In this data subset, we performed a weighted least squares fit and a
bootstrap error analysis on seven measurements of $\dot\nu$. 
This resulted in $n=2.61\pm0.07$. 
We repeated this analysis for the eight measurements of
$\dot\nu$ prior to the magnetar-like
outburst, resulting in $n = 2.68\pm0.03$. The three values of 
$n$ are in agreement 
with each other and with the value obtained from a fully phase-coherent
timing analysis \citep[$n=2.65\pm0.01$;][]{lkgk06}.
All pre-outburst measurements of $n$ are systematically larger than
that measured post-outburst. 
Note that the uncertainties on
the two measurements of $n$ from partially coherent analyses 
(with bootstrap uncertainties) to
pre-outburst data are smaller than the uncertainty for the post-outburst
value of $n$, despite similar data spans fitted in each case. This is
indicative of an increase in the timing noise post-outburst, 
discussed further below. 

A complicating factor in timing some magnetically active
neutron stars is that pulse profile changes often
accompany radiative changes and/or
glitches \citep[e.g.,][]{kgw+03}. This can affect
measured timing parameters and must
therefore be quantified. It has previously been
reported that no significant changes in the pulse
profile were detected during the outburst \citep{lkg10,kh09}.
We further verified that the pulse profile in the $\sim$2 year
period immediately prior to the outburst is not statistically 
different from the summed profile from the $\sim$2 year
of data used to measure $n$, shown in Figure~\ref{fig:kes_normprofile}. 
 
In order to further analyze the timing noise contaminating these data, 
we performed a second partially coherent timing analysis, this time
with each data subset having $\nu$, $\dot\nu$, and $\ddot\nu$ fitted.
The same conditions of ${\chi^2}_\nu \sim 1$ and Gaussian-distributed
residuals, were applied to determine the length of 
each subset, and
once again, the period between 2006 May 31 and 2007 January 27 was
excluded owing to the lack of a coherent timing solution.
Six values of $\ddot\nu$ were obtained before the outburst, while
nine values of $\ddot\nu$ were obtained after the outburst, shown in
Figure~\ref{fig:kes_2der_2008_10}. The measurements of $\ddot\nu$
before the outburst indicate the regular rotation of the pulsar
during this period, while the single value of $\ddot\nu$ above the
average (visible in the inset) occurs
directly after the candidate glitch near MJD~52910,
providing the best evidence for a glitch during this period.
As visible in the figure,
the value of $\ddot\nu$ changed dramatically immediately
after the outburst, to a maximum of $\sim$200 times the quiescent
value, as well as varying in sign, indicating a dramatic increase in
timing noise during the period of glitch recovery. 
The magnitude of $\ddot\nu$ decays
as the glitch recovers during 2007. The inset of
Figure~\ref{fig:kes_2der_2008_10} shows the variation of
$\ddot\nu$ on a smaller scale. While the effects of glitch recovery
and timing noise have subsided substantially by 2008, the post-outburst
variation of $\ddot\nu$ remains larger than its pre-outburst behavior.

\subsection{Timing Noise}
Qualitatively, the timing noise in the 2.2 year period used
to obtain the post-burst measurement of $n$ is larger than
that observed prior to the outburst, though much smaller than in the
initial aftermath of the outburst, when no phase-coherent timing
solution was possible. One measure of the change in timing noise can be found by 
comparing the RMS residuals from MJDs~54492~--~55308             
with those in a similar time span before the outburst.             
Fitting $\nu$, $\dot\nu$, and $\ddot\nu$ phase coherently to the 2.2
year 
segment of data spanning MJD~53086~--~53879, just before the
outburst, results in a timing solution with RMS residuals of 11.4-ms
(0.035~periods), a factor of $\sim$5.5             
smaller than the RMS residuals from MJDs~54492~--~55308,
of 63.6\,ms (0.19~periods). This shows that the
timing noise post-burst is significantly larger 
than before the magnetar-like outburst. This is reflected in the   
new measurement of $n$ by the increased uncertainty               
compared to that on the pre-outburst value               
of $n$ for similar measurement baselines. 

Another measure of the increase in
timing noise is the time span that can be included in each
partially coherent measurement of 
$\dot\nu$ as shown in Figure~\ref{fig:kes75_allnudot}.
When the pulsar is less noisy, more data can be
included in each short timing solution while
satisfying the conditions that ${\chi^2}_\nu \sim 1$
and that no red noise-like structure remain
in the data. 
The pulse profile and pulsed flux are steady, important because
variability in either could cause changes in TOA uncertainties, and
thus affect the time span for each $\dot\nu$ measurement. 
From 2000 until 2006 May, on average,
each measurement of $\dot\nu$ was obtained with data spanning
$111\pm26$\,days, 
while in 2007 (when glitch recovery was still a
significant effect), each measurement of $\dot\nu$
spanned an average of $33\pm 20$\,days. From 
2008 to 2010, the average
time span for each measurement was $68\pm 16$ 
days. Thus, nearly four years post-outburst, the pulsar remains
noisier than prior to the outburst. 

A well known measure of timing noise is the $\Delta_8$
parameter, defined as the contribution to the
rotational phase of the pulsar from a measurement of $\ddot\nu$ over a
period of $10^8$\,s assuming that $\ddot\nu$ is entirely
dominated by timing noise \citep{antt94}.
This parameter is of limited value for a pulsar where
$\ddot\nu$ is dominated instead by secular spin-down, where most of the phase
contribution from $\ddot\nu$ is due to magnetic braking or another
deterministic spin-down mechanism. To quantify the change
in timing noise observed in PSR~J1846$-$0258,
we define a similar parameter which
quantifies the contribution to the rotational phase
from the measurement of the third frequency derivative,
$\nudotdotdot$, over a time span of $\sim 2.5\times 10^{7}$\,s.
The time span is optimized for this
particular pulsar, as it is the approximate amount of time required
to obtain a significant measurement of $\nudotdotdot$,
while allowing several measurements to be made given the available
data span. Thus, in
analogy with the $\Delta_8$ parameter we define: 
\begin{equation}
\Delta_{\nudotdotdot} \equiv {\rm{log}} \left ( \frac{1}{24}
\frac{|\nudotdotdot|(2.5\times10^7)^4}{\nu} \right). 
\end{equation}
We measured the $\Delta_{\nudotdotdot}$ parameter for each
$\sim 2.5\times10^7$\,s segment of data where a phase-coherent
timing solution was available, and have shown the results in 
Figure~\ref{fig:kesnoise}. The value of $\Delta_{\nudotdotdot}$
increases dramatically after the 2006 outburst, after which it decays,
but by 2010
has not returned to the pre-outburst quiescent level. 

\section{CHANDRA Observations and Analysis}
\psr\ was previously observed with the \emph{Chandra X-ray
Observatory} in 2000 and 2006 with 37-ks and 155-ks exposures,
respectively. Serendipitously, the latter observation was carried
out seven days after the outburst. We obtained a new \emph{Chandra} 
imaging observation with a 44.6-ks ACIS-S exposure (ObsID 10938) 
on MJD~55053 (2009 August
10), over three years after the outburst. The observation was taken 
in 1/8 subarray mode, which gives a short frame
time of 0.4-s, resulting in negligible ($<3\%$) pileup of the
pulsar counts. All the data reduction and analysis were performed
with CIAO~4.1\footnote{\url{http://cxc.harvard.edu/ciao4.1/}}.
Figure~\ref{fig:chandra} shows the exposure-corrected images
of \psr\ and the associated PWN in the 
1~--~7\,keV energy range. While the PWN
exhibits no obvious change in the overall morphology between
the three observations, time variabilities are observed in some
small-scale features.  
The most prominent area of variability is the northern
clump located $\sim7\arcsec$ northeast of the pulsar.
We extracted the count profiles of the clump from the
exposure-corrected images using 6$\arcsec$ wide boxes
(as indicated in Figure~\ref{fig:chandra}), and
show the results for each epoch in Figure~\ref{fig:clump}. The clump morphology
evolved from a single peak in 2000 to a double peak in 2006,
and back to a fainter single peak in 2009. However, there
is no evidence for bulk motion of the clump, 
as is visible in the figure. 

For the spectral analysis, we extracted the spectrum of the entire
PWN excluding the central 3\arcsec\ radius, and fitted it with
an absorbed power-law model. The column density was fixed at
$N_{\rm{H}}=4.0\times10^{22}$\,cm$^{-2}$ during the fit, to
provide a direct comparison with previous studies \citep{nsgh08,ks08}. 
In the 2009 observation, we found a
photon index of $\Gamma=1.90\pm0.03$ and an absorbed flux of 
$f^{\rm abs}_{0.5-10}=(1.37\pm0.05)\times10^{-11}$\,erg\,s$^{-1}$\,cm$^{-2}$ 
in the 0.5~--~10\,keV range (hereafter, quoted uncertainties
are 90\% confidence levels). This is fully consistent with the
PWN spectrum in 2000, which has $\Gamma=1.88\pm0.03$ and
$f^{\rm abs}_{0.5-10}=(1.39\pm0.02)\times10^{-11}$\,erg\,s$^{-1}$\,cm$^{-2}$
\citep{nsgh08}. 

For the pulsar itself, counts were extracted from a 2\arcsec radius
aperture and grouped with 70 counts in each spectral bins.
We followed Ng et al.\ (2008) to account for
the nebular contamination using a power-law model with fixed
$\Gamma=1.9$ and $f^{\rm abs}_{0.5-10}=1.0\times
10^{-12}$\,erg\,s$^{-1}$\,cm$^{-2}$.
As in previous studies, we started with a power-law plus blackbody
model, but found that the latter component is only marginally
detected, statistically insignificant. The pulsar spectrum in 2009
is adequately fitted by an absorbed power-law model with
$\Gamma=1.1\pm0.1$, and we derive an upper limit on the
blackbody temperature of 0.25\,keV at the 90\% confidence
level, significantly lower than the value 0.9$\pm$0.2\,keV measured
during the 2006 outburst. Table~\ref{table:chandra},
compares the best-fit spectral parameters among the three epochs,
indicating that the pulsar flux in 2009 is at a similar level as
in 2000, much lower than that in 2006.
For completeness, we also fitted a two blackbody model to the 2009
{\emph{Chandra}} observation. While the goodness of fit 
is statistically similar to that obtained when fitting a
power-law model, the two-blackbody model gives an unphysically high
temperature of $1.6\pm0.1$\,keV for the hotter thermal component.
In addition, there is evidence for a continuous power-law component
from 1 to 300\,keV \citep{kh09}, rendering the two blackbody model of
limited interest for this pulsar.

\section{Discussion}
\label{sec:discussion}

\subsection{Timing Noise and the Braking Index}

We have observed a decrease in the braking index of \psr\
from $n=2.65\pm0.01$ to
$n=2.16\pm0.13$, corresponding to $\Delta{n} = -0.49\pm0.13$, or a
decrease of $18\%\pm5\%$. The change in $n$ was accompanied
by an increase in the timing noise of the pulsar, which remains larger
than the pre-outburst level, nearly four years after the glitch
and outburst on 2006 May 31. Previous long-term observations of
$n$ in young pulsars have shown
that they are remarkably steady. In PSR~B1509$-$58, timing observations over
21 years show that $n$ varies by only $\sim 1.5\%$ \citep{lkgm05}, while
the Crab pulsar exhibits variations on the order of $5\%$ \citep{lps93}.

There are two possible ways to interpret the measurement
of $\Delta{n} = -0.49\pm0.13$. The first
is that the true $n$ decreased by a significant amount after the
2006 outburst. The second
is that the increased timing noise
is causing an apparent decrease in $n$.
We discuss each of these interpretations next.

\subsubsection{Variable Braking Index} 

If the true value of the braking index changed permanently 
at the time of the
magnetar-like outburst, what could be the physical cause of such an
effect? From the spin-down law derived assuming magnetic dipole
braking \citep[where $I$ is the moment of inertia, $R$ is the
neutron star radius, and $\alpha$ is the angle between the spin and
magnetic axes; e.g.,][]{go69}, 
\begin{equation}
\dot \nu = \frac{-8\pi^2}{3} \frac{{B}^2 {R}^6{\rm{sin}}^2{\alpha} }{Ic^3} \nu^3,
\label{eqn:spindown}
\end{equation}
we can infer that $d^2I/dt^2 > 0$\footnote{Note that the different
relationship between $I$ and $\Delta{n}$ than for $B$ and $\alpha$
arises from correctly including the variability of $I$ in the spin-down
luminosity, $\dot{E}$,  when deriving the above spin-down law. The full form of the
spin-down luminosity is $\dot{E} = 4{\pi}I{\nu}{\dot\nu} +
4{\pi}{\nu}^2 {\dot{I}}$, where only the first term is considered 
if $I$ is constant. Using the full form of
$\dot{E}$ in the derivation of the spin-down law leads to the
dependence of $d^2I/dt^2$ on $\Delta{n}$.}, $d\alpha/dt>0$, or  $dB/dt>0$ will 
cause $n<3$. 

\begin{figure}
\plotone{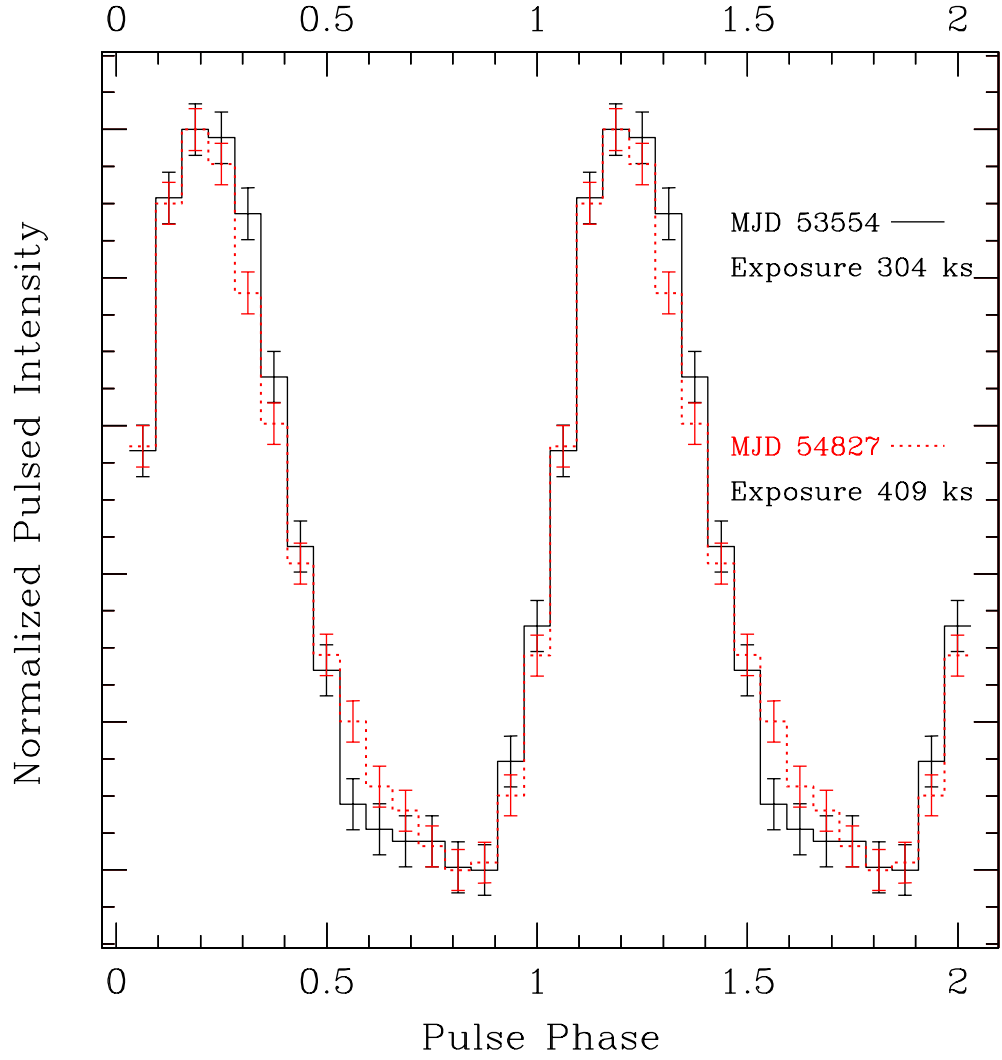}
\figcaption[Normalized pulse profiles]{\footnotesize{
Pulse profile of PSR J1846$-$0258 for two
sections of data of $\sim$2-years. Two cycles are shown for clarity. 
The solid line shows the pulse profile before the  outburst
in 2006 May. The dotted (red in the online version) line shows the summed
pulse profile from 2008 January to 2010 April. Subtracting the two
profiles results in residuals with ${\chi^2}_\nu \sim 1.3$,
where the probability of this ${\chi^2}_\nu$ or higher
occurring by chance is 19\%. Thus, the two profiles are not
statistically significantly different.}
\label{fig:kes_normprofile} }
\end{figure}

\begin{figure}
\plotone{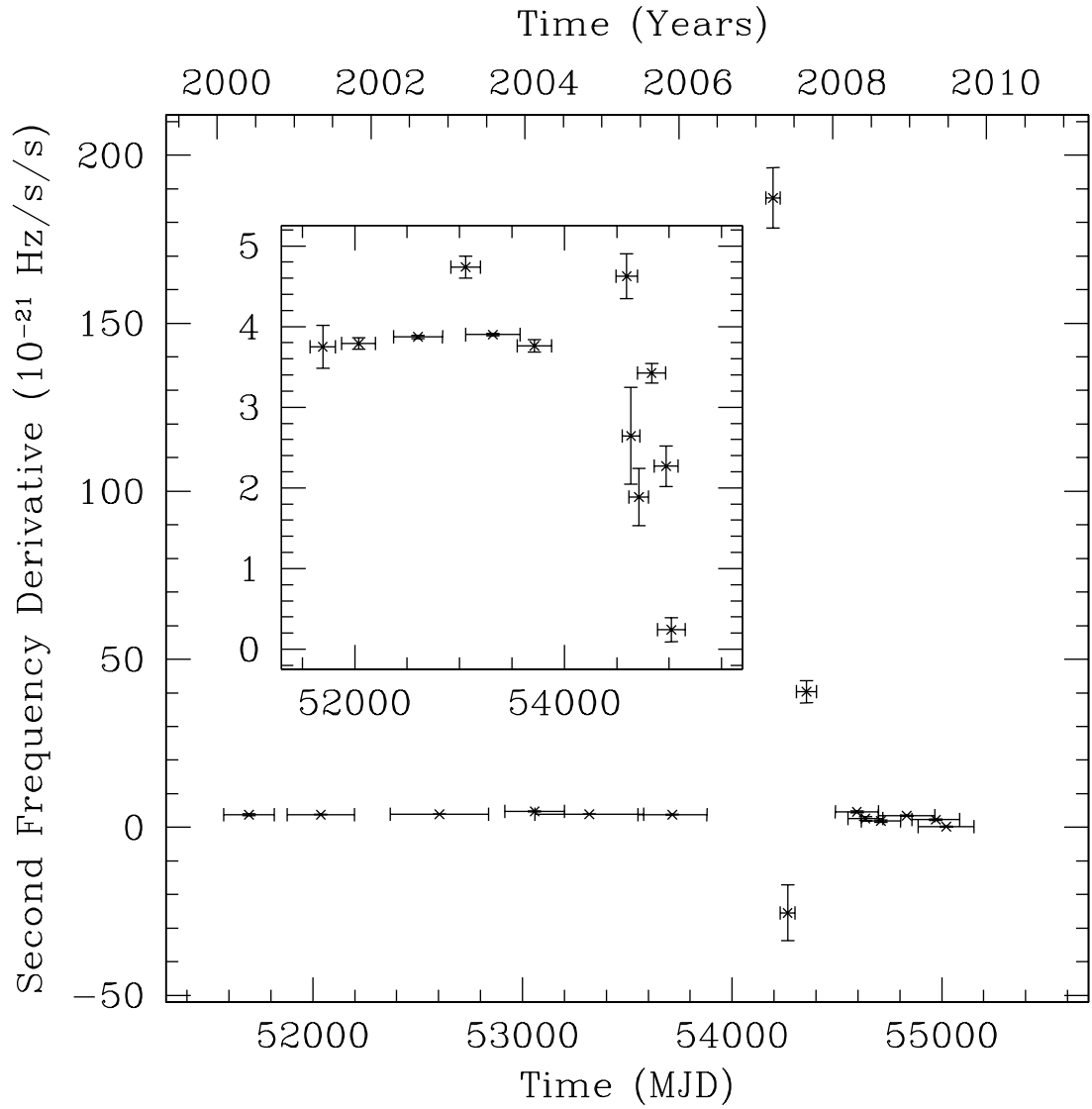}
\figcaption[Second frequency derivative measurements for \kes.]
{\footnotesize{Second derivative measurements for \kes\ from 2000 to 2010.
The three measurements occurring after the glitch and outburst are one
to two orders of magnitude larger than the subsequent measurements and vary
in sign, indicating that they are severely contaminated by glitch
recovery and/or timing noise. The inset shows measurements of
$\ddot\nu$ on a smaller scale to highlight the smaller variation away
from the glitch recovery. The variation in $\ddot\nu$ during the
period from 2008 to 2010 is
larger than in the pre-outburst era, and the mean value is 
systematically smaller. The
one value of $\ddot\nu$ pre-burst that is significantly larger than
the average is directly after the candidate glitch near
MJD~52910, and is the best evidence that a glitch actually occurred
at that epoch.}
\label{fig:kes_2der_2008_10} }
\end{figure}

\begin{figure}
\plotone{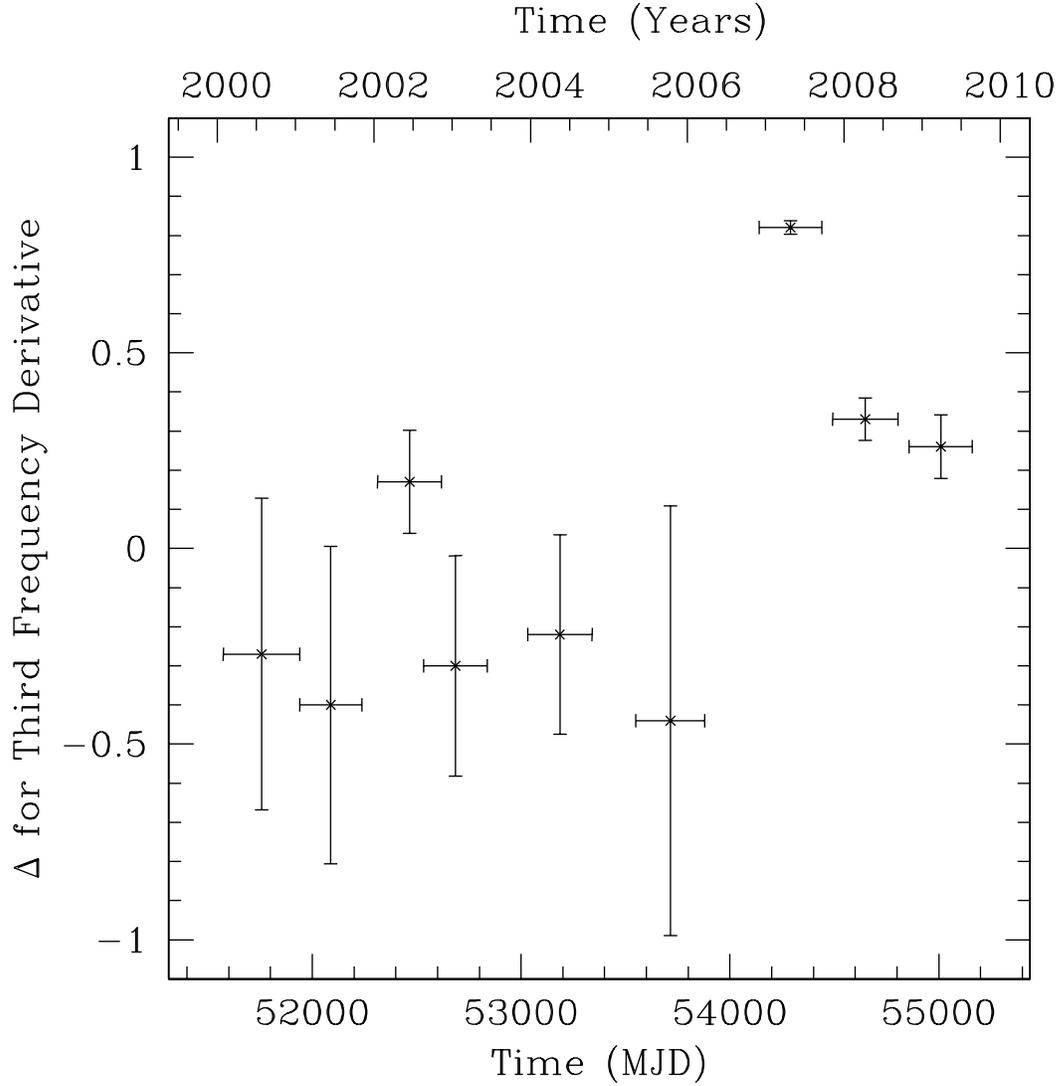}
\figcaption[$\Delta_{\nudotdotdot}$ parameter for \kes.]{\footnotesize{
Quantification of the timing noise in \kes\ over 10-year.
Each point is a measurement of the $\Delta_{\nudotdotdot}$ parameter
for approximately 2.5$\times 10^7$\,s. This provides an estimate
of the amount of timing noise observed in the pulsar, and shows a
dramatic increase after the large glitch observed in 2006. }
\label{fig:kesnoise} }
\end{figure}

\begin{center}
\begin{figure}
\plotone{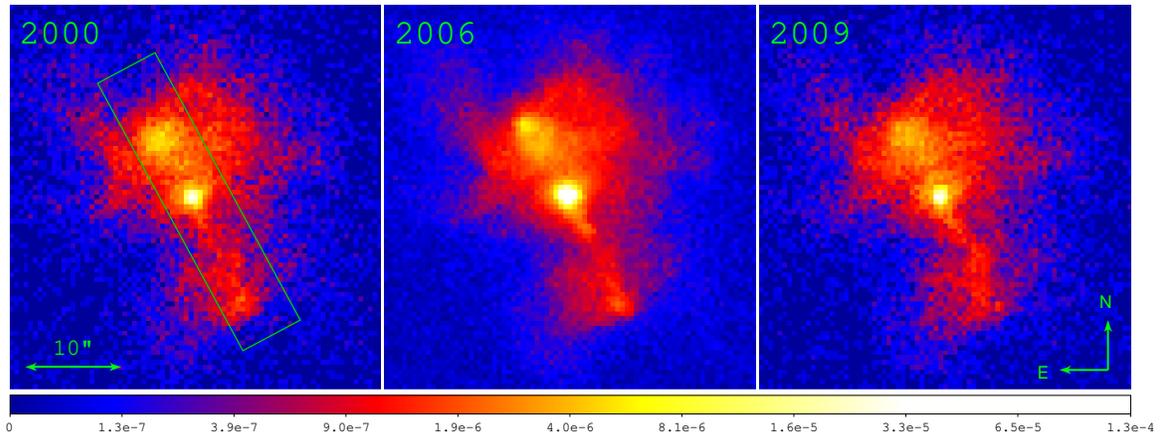}
\figcaption[Chandra]{\footnotesize{Exposure-corrected {\textit{Chandra}} ACIS 
images of the PWN associated with \psr\ in 1~--~7 keV.
The $6''$ wide box along a position angle of 27$\degrees$ (north through
east) is analyzed in detail for each epoch in Figure~\ref{fig:clump}.}
\label{fig:chandra} }
\end{figure}
\end{center}

\begin{figure}
\plotone{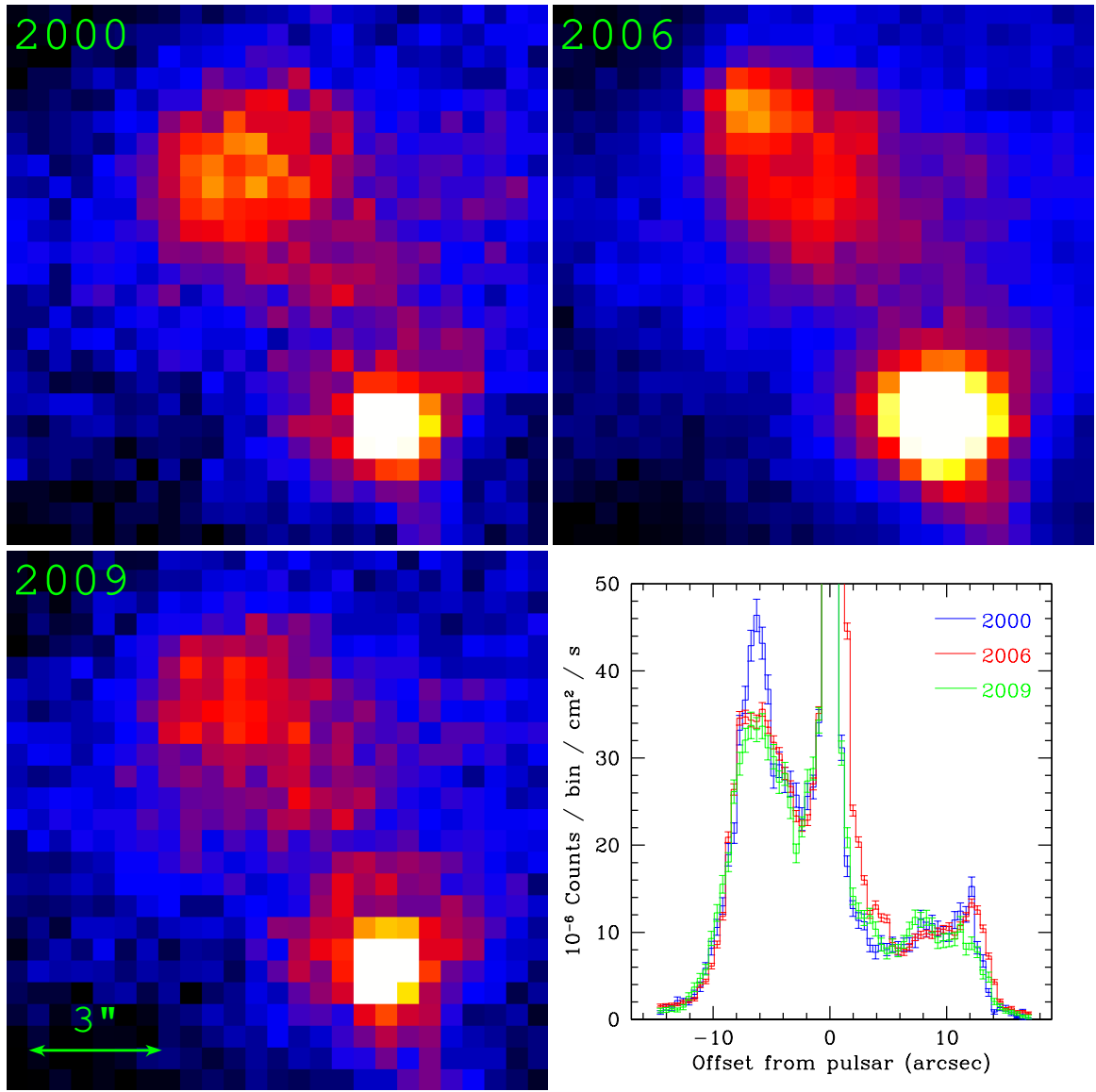}
\figcaption[clump]{\footnotesize{
Zoom-in of Figure~\ref{fig:chandra} showing the
northern clump of the PWN (boxed region in Figure~\ref{fig:chandra}) 
in the 1~--~7 keV range. The bottom right panel       
illustrates the count profiles extracted from the $6\arcsec$ wide box
(as shown in Figure~\ref{fig:chandra}),
indicating the evolution of the clump between epochs.}
\label{fig:clump}}                 
\end{figure}

While it is hard to imagine a physical situation causing an
accelerated growth or decay in $I$, varying values of $\alpha$ or $B$
have been considered in the
past. Counter-alignment of the magnetic field (i.e., an increasing $\alpha$)
results in $n<3$, even if $\Delta{n}=0$. A
sudden increase in the rate of change of $\alpha$ would produce
$\Delta{n}<0$. However, this is difficult to
invoke for the observations of \kes\ because the pulse
profile shows no variation over the relevant time
period (e.g., Figure~\ref{fig:kes_normprofile}). A small 
change in $\alpha$ could still be possible if our line of
sight is crossing the center 
of the pulsar beam, as a large change in the pulse profile may
not be required from a small change in field orientation.
However, this is hard to reconcile with the lack of detected
radio pulsations from the source \citep{aklm08}, which is
typically interpreted as our line of sight missing the magnetic
pole, from where the radio emission is thought to originate, and
crossing only the wider X-ray beam. 
Alternatively, radio emission from \psr\ may be
suppressed as a result of its large magnetic field, as suggested 
for most known magnetars \citep{bh98}, although radio emission has
been detected from several pulsars with higher inferred magnetic
dipole strengths than observed from \psr. Nevertheless, if radio 
emission were suppressed in
\psr, then our line of sight may cross the center of the X-ray beam, 
allowing for a small change in $\alpha$ without accompanying profile
changes. This point cannot be resolved, however, without further
details of the geometry of the system.  

An increase in the magnitude of $B$, without 
a change in the orientation
of the field could also cause a decrease in $n$ \citep{br88,bah83}.
Making all the assumptions of perfect dipole spin-down but             
allowing $dB/dt >0$, a braking index of $n=2.65$ implies a             
time scale of growth for the magnetic field of $\sim$8000-year,
while $n=2.16$ shortens the growth timescale to $\sim$3500-year.      
The possibility of magnetic field growth is intriguing given 
the magnetic activity observed prior to the change in $n$.
\citet{bs07} suggest that magnetar-strength fields emerge over a period of time, as
shielding currents dissipate. If the effective $B$ is currently in 
such a period of growth, the smaller value of $n$ could result
\citep{lyn04}. However, this picture lacks a description of why the
$B$-growth timescale should have increased so rapidly. 

The standard spin-down law (Equation~\ref{eqn:spindown}) is a major
idealization even for rotation-powered pulsars; for magnetars, the
picture is almost certainly much more complicated.
According to one version of the magnetar model, there is a one-to-one
relationship between $n$ and the large-scale
twist angle between the north and south hemispheres of magnetic field,
$\Delta\phi_{\rm{N-S}}$ \citep{tlk02}. Here, a decrease
in $n$ would imply an increase in the twist angle.
A braking index of $n=2.65$ corresponds to a twist angle of
$\Delta\phi \simeq 1$\,rad, while $n=2.16$ implies a twist angle of
$\Delta\phi \simeq 2$\,rad. 
If the magnetic field remains approximately constant before and after
the outburst, such an increase in the twist angle should be accompanied 
by an increase in
the X-ray luminosity of $\sim50\%$, whereas no persistent increase in
$L_{\rm{X}}$ is observed in \psr\ as evidenced by the consistent flux value
measured with {\textit{Chandra}} in 2000 and 2009. 
Furthermore, while the above model assumes a global magnetic field twist, 
\citet{bel09} argues that such global self-similar twists do not present a
viable explanation 
for magnetar behavior. Instead, he suggests that magnetospheric
currents are confined to narrow regions on the most extended field
lines, which are not responsible for the bulk of magnetar X-ray emission. 
If this picture is correct, the spin-down of the star can 
vary without accompanying changes in X-ray luminosity.
In addition, this model predicts that 
an increase in spin-down torque (though not necessarily monotonic)
should occur sometime 
after a radiative event, and should eventually return to the
pre-outburst torque value. Qualitatively, this seems to present an
explanation of the observed timing variability in \psr; 
however, it provides no quantitative prediction for $n$. 
A variable $n$ provides an unambiguous test of the theory of spin-down
presented by \citet{mel97}. He posits that the radius relevant to
neutron star spin-down is not the point-like neutron star radius, but
a somewhat larger ``vacuum radius'' where field aligned flow breaks
down. This radius is large enough that the system can no longer be
treated as a point dipole, resulting in modifications to the standard
spin-down predictions. In the context of this model, a measurement
of $\nu$, $\dot\nu$, and $\alpha$ uniquely predicts $n$. While there
is no estimate of $\alpha$ for \psr, there are now two
measurements of $n$ so that the theory can be tested. Given the observed
change in $n$, and assuming that $\alpha$ is stable over the
magnetar-like outburst, this theory predicts that the magnetic field
should have increased by a factor of $\sim$6. This is not observed,
however, as the magnetic field estimate has increased by just 0.3\%
compared with the pre-outburst value, in contradiction to this theory.
Alternatively, if $B$ is roughly constant as observed, and instead
 $\alpha$ were allowed to change at the time of the
event, \citet{mel97} predicts that the angle between the
spin and magnetic axes would have changed from $\sim$9$\degrees$
to $\sim$4$\degrees$. Such a change might be visible as differences in 
the pulse profile for some lines of sight, but cannot be excluded
given the lack of observed profile changes.

Another possible explanation put forth for a static measurement of
$n<3$ is that magnetic field lines are
deformed due to plasma in the magnetosphere \citep{br88}. 
A sudden increase in the amount of plasma in the magnetosphere of
\kes, perhaps injected at the time of the magnetar-like outburst,
could cause $n$ to decrease. The best evidence for a plasma-filled 
magnetosphere affecting pulsar
spin-down is found from the ``intermittent'' radio pulsar, PSR~B1931+24,
which has dramatic, quasi-periodic changes in $\dot\nu$ correlated
with radio pulsations which turn on and off \citep{klo+06}.
\cite{hck99} propose that spin down can arise from a 
combination of magnetic dipole radiation and wind
losses. An increase in losses from the wind
relative to dipole radiation will manifest as a smaller value of $n$. 
The spin-down formula given by \citet{hck99} implies a braking index of 
\begin{equation}
n = 3 - \frac{2\nu}{{\dot\nu}} \frac{ {L_p}^{1/2}B R^3}{2I
\sqrt{6c^3}}, 
\end{equation}
where $L_p$ is the kinematic luminosity of the wind, which can, in turn, be
estimated with a measurement of $n$. 
For \psr, a change in $n$ from 2.65 to 2.16 corresponds to 
nearly an order of magnitude increase in the persistent
particle luminosity from $L_p\simeq 1\times10^{36}$\,erg\,s$^{-1}$ to
$L_p \simeq 6\times 10^{36}$\,erg\,s$^{-1}$. 
Because the outflowing particles travel near the speed of light, the
additional particles would populate the PWN and be
observable as a factor of $\sim$6 increase in the PWN flux in the
2009 {\emph{Chandra}} observation as compared to the observation in
2000. However, no such flux increase is detected (see Section 4). 
The lack of a PWN flux increase seems to refute the idea that
an increase in wind losses is responsible for $\Delta{n}<0$, however,
a more rigorous derivation 
of the relationship between wind losses and
spin down may provide further insight into this issue. 

Interestingly, because a change in the plasma conditions in the
magnetosphere might also affect magnetospheric torques, 
this could possibly explain an increase in timing
noise \citep[e.g.,][]{che87a}. As shown by \citet{lhk+10},
timing noise in some pulsars can be traced to magnetospheric
fluctuations. For these radio pulsars, there is a correlation between
torque variations and pulse shape changes.
This is difficult to apply in the case of \psr\ because 
no radio pulsations
have been detected \citep{aklm08}, and there is no evidence
for profile variability in the X-ray band (see
Figure~\ref{fig:kes_normprofile}).

\citet{cs06} note that if the co-rotation radius of the magnetosphere,
$r_{\rm{co\mbox{-}rot}}$, is less than the light cylinder radius,
$R_{\rm{LC}}$, owing to imperfect reconnection within the
magnetosphere, $1\le n \le 3$ will result. If the co-rotation
radius decreased at the time of the magnetar-like outburst from
magnetic field lines opening, $n$ would also decrease. 
They parameterize the relationship between 
$r_{\rm{co\mbox{-}rot}}$ and $n$ as
\begin{equation}
r_{\rm{co\mbox{-}rot}} = R_{\rm{LC}}  \left (\frac{\nu}{\nu_0}\right)^\frac{3-n}{2}.  
\end{equation}
If the initial spin-period of \psr\ was $P_0=1$\,ms, the
implied co-rotation radius prior to outburst was
$r_{\rm{co\mbox{-}rot}}= 0.36R_{\rm{LC}}$ when $n=2.65\pm0.01$, and
would have decreased to 0.09$R_{\rm{LC}}$ post-outburst when $n=2.16\pm0.13$.
Changes to the extent of the co-rotation region of the magnetosphere
might be visible as changes in the pulse profile; however, no such
changes have been observed from \psr. In the context of this model,
however, it is impossible to rule out changes to the extent of the 
magnetosphere as 
no independent estimate of the initial spin period is available. 
In the context of this model, taking $P_0=1$\,ms as the lower limit on
the birth spin period for \psr,  
provides a lower limit on the co-rotation
radius for any measured value of $n$. 
Furthermore, $P_0=1$\,ms provides an upper limit on the change in
co-rotation radius at the two epochs, since this quantity decreases
with slower birth spin periods.

\subsubsection{Timing Noise}
The second possible interpretation of $\Delta{n}<0$ is that the 
true $n$ is constant but remains masked by the high timing noise. 
The increase in timing noise could arise as a result of changes to the
superfluid interior brought on by the unusual 2006 glitch or changes
in the magnetosphere after the outburst.  Though a bootstrap error analysis
was employed in order to better account for the effect of the increase
in timing noise, a definitive test is not possible. 
Continued timing observations may be able to solve this issue,
if the timing noise level continues to decrease and the new
braking index remains steady. However, it is also possible that
the increased level of timing noise and the decreased braking
index are connected, for example, via an increase in the
magnetospheric plasma density. In that case, if the pre-outburst 
conditions are eventually reobtained, both the timing noise and
braking index should relax to their pre-outburst values, rendering a
temporary value of $n<2.65$ ambiguous in nature. 

Fluctuation in pinned superfluid in the pulsar interior is one of the
possible causes of timing noise \citep[e.g.,][]{anp86}. Thus, one
possible explanation for the increase in timing noise is that
significant changes were imparted to the neutron star interior at the
time of the 2006 glitch and outburst. 
The glitch was followed by an unusual over-shoot recovery ($Q\simeq 8.7$
on a timescale of $\tau_d\simeq$127\,days) and a permanent increase in the
magnitude of $\dot\nu$ 
\citep[with fractional magnitude $\Delta{\dot\nu}/{\dot\nu} \simeq
0.0041$;][]{lkg10}. 
This recovery is thus far unique among pulsar glitches 
and the origin and long-term consequences of such
behavior are not well understood. 
By contrast, the permanent change in $\dot\nu$ following the glitch
is not unusual when compared to those measured after other glitches. 
In addition, we note that 
the change in $\dot\nu$ is not responsible for the observed decrease
in $n$. Because the fractional increase in $\dot\nu$ is
three orders of magnitude smaller than the fractional change in
$\ddot\nu$, it is the latter effect that dominates the change in $n$. 

In addition to the detected glitch recovery and permanent increase in $\dot\nu$, 
non-monotonic variations in $\dot\nu$  were observed in the aftermath of the glitch. 
While glitch recovery (i.e., a temporary increase in $\dot\nu$) and
discrete jumps in $\dot\nu$ accompanying a glitch are both established
phenomena, 
to our knowledge, no other RPP has experienced 
changes in timing noise similar to those observed from \psr. 
However, variable spin-down torque 
has been observed in several magnetars. 
The AXP~1E~1048.1$-$5937 twice experienced approximately year-long
periods of rapid $\dot\nu$ variations, i.e., sudden, but temporary
increases in timing noise \citep{gk04,dkg09}. Similar variations in
$\dot\nu$ were observed in the transient AXP XTE~1810$-$197 after its
2003 outburst \citep{ccr+07}. Thus, the observed change 
in timing noise in \kes\ can be interpreted as yet another
example of magnetar-like properties from this RPP,
even if the phenomenon is currently unexplained. 

\subsection{Phase-averaged pulsar flux and the pulsar wind nebula}
\label{sec:pwndisc}
The \emph{Chandra} results show that the phase-averaged
pulsar flux and spectrum in 2009 have returned to the quiescent
values observed in 2000. 
This agrees with the pulsed flux history from \rxte, where the initial
flux increase in 2006 was observed to decay exponentially with $1/e
\sim 55$\,days \citep{ggg+08}, and has since remained at the quiescent 
level \citep{lkg10}. 
Given that the 2009 \emph{Chandra} data
were taken over 1000 days after the 2006 outburst, our findings are 
not unexpected. The time variability of the small-scale features in the
PWN could be attributed to magnetohydrodynamic instabilities
in the flow, similar to what has been observed in the PWNe
powered by the Vela pulsar and PSR B1509$-$58 \citep{ptks03,dgap06}
and is unlikely to be related to
a sudden deposition of particles at the time of the 
outburst. \citet{ggg+08} reported an energy release
of $\sim 5\times10^{41} (d/6\,{\rm{kpc}})^2$\,erg (2~--~60\,keV)
in the 2006 outburst. In comparison, the
total energy released in the giant flare from the magnetar SGR
1806$-$20 in 2004 is $2\times 10^{46}$\,erg \citep{pbg+05}, 
of which $4\times 10^{43}$\,erg went into particle energy
\citep{gkg+05}. 
The resulting ratio between the particle energy and electromagnetic
radiation is $2\times10^{-3}$. 
If the same ratio
holds for \psr, then the energy of the injected particles
would be only $\sim 10^{39}$\,erg. With a $B$-field strength
of 15$\mu$\,G in the PWN (Djannati-Ata\"{i} et al. 2007)\nocite{ddt+07},
the synchrotron cooling timescale of a particle emitting at 5-keV is
$\sim$300-year. These particles would have induced an X-ray flux
enhancement in the PWN of $2\times 10^{-17}\,$erg\,s$^{-1}$\,cm$^{-2}$. This is
five orders of magnitude lower than the X-ray flux of the northern clump,
and six orders lower than that of the entire PWN, too small to be detected.
Moreover, if the particles travel isotropically with a typical 
post-shock flow speed of
$c/3$, the flux would have already spread 
over the entire PWN of 10\arcsec\ radius. Therefore, we do not 
expect the 2006 outburst to have
any observable effects on the PWN.

\section{Conclusions}

The observed change in $n$ after the magnetar-like outburst in \kes,
if shown to be steady via ongoing timing observations, 
has important implications for the physics of
neutron star spin-down. A decrease in $n$ could have several origins,
and a detailed theoretical framework is necessary for interpreting
this observation. 

Most theoretical descriptions of a changing $n$ require an
accompanying persistent change in radiative behavior of the pulsar,
while we observe neither pulse profile variability or persistent flux
enhancement. An increase in particle wind losses relative to dipole
losses does not provide a good description of $\Delta{n}<0$ for \kes\
because of the lack of a persistent increase in PWN luminosity
\citep{hck99}. However, variability in magnetospheric plasma remains a
promising avenue for future consideration, especially considering the
recent report of variable spin-down rate correlated 
with radio pulse shape changes for several pulsars,
which confirms a link between torque and emission properties in
several pulsars \citep{lhk+10}. No variability in the X-ray pulse
profile is detected in \kes; however, small shape changes may be
present but not observable in current data. In addition, short
time-scale variability would not be detectable in \rxte\ 
observations which are typically from 1.5 to 2\,hr long. 

The timing noise in \psr\ is observed to be of a higher level than 
prior to the outburst. That is, even four years after the glitch
and magnetic activity, the pulsar is rotating less regularly than in
its pre-outburst quiescent state. 
It is interesting to note, however, that the current 
timing noise in \psr\ would not be unusual if observed in any young 
pulsar. Rather, it is the sudden change in
the level of timing noise in \psr\ that is noteworthy. Since the timing 
noise is simply of a higher level and not otherwise different from that
observed in other pulsars, the phenomenon cannot be used as a diagnostic
of previous unseen magnetic activity in other pulsars. 

The observed decrease in $n$ and increase in timing noise             
reported here may or may not be permanent. Regular             
monitoring observations beyond the \xte\ era may help to             
answer this question, as well as to search for future             
magnetar-like X-ray outbursts and glitches from \kes.  

The 2009 {\emph{Chandra}} observations of \kes\ show that the total 
flux and spectrum are consistent with the quiescent values observed
in 2000. No significant variability was detected in the PWN that can be
associated with the 2006 outburst, nor, given the energetics of the
outburst, would any be expected. 
The variability of the PWN observed in the 2006 {\emph{Chandra}}
observation is most likely unrelated to the outburst, and is
probably similar in origin to the variation in small-scale 
features seen in other PWNe. 

\acknowledgments
We thank A. Beloborodov, A. Melatos, and an anonymous referee for
comments that improved the manuscript. 
This research made use of data obtained from the High Energy Astrophysics
Science Archive Research Center Online Service, provided by the NASA-Goddard
Space Flight Center. CYN is a CRAQ postdoctoral fellow. 
VMK holds the Lorne Trottier Chair in
Astrophysics and Cosmology and a Canada Research
Chair in Observational Astrophysics. Funding for this work was 
provided by NSERC Discovery Grant Rgpin 
228738-03, FQRNT, and CIFAR. 


\clearpage

\centering

\clearpage

\clearpage


\begin{table}[tbp]
\centering
\caption{2008--2010 Spin Parmeters for \psr\ \label{table:kes75_3_parameters}}
\begin{tabular}{lc}
\hline
{{Phase-coherent timing analysis}} & \\
\hline \hline
Date range (Modified Julian Day)   & 54492.089~--~55308.598\\
Date range (Years)                 & 2008 Jan 27~--~2010 Apr 22\\
Number of TOAs                     & 100 \\
Epoch (Year)                       & 2009 Mar 1\\
Epoch (MJD)                        & 54834.0\\
$\nu$ (Hz)                         & 3.0621185502(4) \\
$\dot\nu$ ($10^{-11}$~s$^{-2}$)    & $-$6.664350(2) \\
$\ddot{\nu}$ ($10^{-21}$~s$^{-3}$) & 2.725(3) \\
Number of derivatives fitted       & 2 \\
RMS residuals (ms)                 & 63.6\\
\hline
{{Partially phase-coherent timing analysis}}  &\\
\hline \hline
$\ddot{\nu}$ ($10^{-21}$~s$^{-3}$) & 3.13(19)\\ 
Braking index ($n$)                & 2.16(13)\\
\hline
\end{tabular}     \\
\footnotesize{Quoted uncertainties are the formal $1 \sigma$
uncertainties as repored by
TEMPO \\
for the fully phase-coherent timing analysis.
Details about the uncertainty on $\ddot\nu$ and \\
$n$ for 
the partially coherent timing solution are from a bootstrap 
analysis, as explained 
in the text (see Section~3.1).}
\end{table}


\begin{table}[tbp]
\centering
\caption{Variation of $n$ with Number of Fitted Frequency
Derivatives \label{table:n_v_der} }
\begin{tabular}{lcc}
\hline
{{Derivatives}} &{{Braking Index}}  & ${\chi^2_\nu}$ \\
{{fitted}}      & ${n}$     &                \\
\hline \hline
2 & 1.888(2) & 39.71  \\
3 & 2.010(3) &  6.12 \\
4 & 1.980(8) &  6.00 \\
5 & 2.05(1)  &  5.48 \\
6 & 2.08(2)  &  5.50 \\
7 & 2.51(4)  &  3.51 \\
8 & 2.60(4)  &  3.26 \\
9 & 2.95(7)  &  2.93 \\
10 & 2.91(7) &  2.73 \\
11 & 2.1(1)  &  2.12 \\
12 & 2.1(1)  &  2.14 \\
\hline
\end{tabular} \\
\footnotesize{Braking index variation with number
of fitted derivatives. The number \\
of degrees of 
freedom for two fitted
derivatives is 96. Uncertainties are the \\
formal 
1$\sigma$ uncertainties 
returned by TEMPO for the timing solution \\ spanning 
MJD~54492~--~55308. }
\end{table}


\begin{table}[tbp]
\centering
\caption{SPECTRAL FITS TO PSR J1846$-$0258\label{table:chandra}}
\begin{tabular}{ccccccccc}
\hline
Epoch &$N_{\rm H}$           &$\Gamma$&$f_{0.5-10}^{\rm abs,PL}$&$f_{0.5-10}^{\rm unabs,PL}$&$k$T &$f_{0.5-10}^{\rm abs,BB}$&$f_{0.5-10}^{\rm unabs,BB}$&$\chi^2_\nu$/dof\\
      &($10^{22}$\,cm$^{-2}$)&        &{($10^{-12}$\,erg}       &{($10^{-12}$\,erg}         &(keV)&{($10^{-12}$\,erg}       &{($10^{-12}$\,erg}             &       \\
      &                      &        &{cm$^{-2}$\,s$^{-1}$)}   &{cm$^{-2}$\,s$^{-1}$)}     &     &{cm$^{-2}$\,s$^{-1}$)}   &{cm$^{-2}$\,s$^{-1}$)}         &       \\
\hline \hline
{2000} & $4.0^\dagger$       &$1.0^{+0.8}_{-0.3}$ & $4.2\pm0.2$ & $6.1\pm0.4$ & $<0.8$ & $<0.04$ & $<0.4$ & 0.94/38 \\
{2006} & $4.0^\dagger$       &$1.9\pm0.1$& $1.3\pm0.3$ & $31\pm6$ & $0.9\pm0.2$ & $1.7\pm0.2$ & $3.2\pm0.4$ & 0.99/134\\
{2009} & $4.0^\dagger$       &$1.0\pm0.1$ & $3.6\pm0.5$ & $5.2\pm0.7$ & $<0.25$ & $<0.015$ & $<0.45$ & 0.84/72 \\
\hline
\end{tabular} \\
\footnotesize{
{\dag}{ -- held fixed in the fit.} \\ 
The uncertainties quoted are 90\% confidence intervals
and upper limits are at the 90\% confidence level. \\
The 2000 and 2006 results are from Ng et al.\ (2008).}
\end{table}

\end{document}